\documentclass[a4paper,11pt]{article}

\usepackage{jcappub}

\usepackage{amsfonts,graphics,epsfig,subfigure}

\title{A consistent and unified picture for critical phenomena of f(R) AdS black holes}

\author[a,b,1]{Jie-Xiong Mo,\note{Corresponding author}}
\author[a,b]{Gu-Qiang Li,}
\author[b]{Yu-Cheng Wu}

 \affiliation[a]{Institute of Theoretical Physics, Lingnan Normal University, Zhanjiang, 524048, Guangdong, China}
\affiliation[b]{Department of Physics, Lingnan Normal University, Zhanjiang, 524048, Guangdong, China}

\emailAdd{mojiexiong@gmail.com}
\emailAdd{zsgqli@hotmail.com}
\emailAdd{wuyucheng0827@163.com}

\abstract{A consistent and unified picture for critical phenomena of charged AdS black holes in $f(R)$ gravity is drawn in this paper. Firstly, we investigate the phase transition in canonical ensemble. We derive the explicit solutions corresponding to the divergence of $C_Q$. The two solutions merge into one when the condition $Q_c=\sqrt{\frac{-1}{3R_0}}$ is satisfied. The curve of specific heat for $Q<Q_c$ has two divergent points and can be divided into three regions. Both the large radius region and the small radius region are thermodynamically stable with positive specific heat while the medium radius region is unstable with negative specific heat. However, when $Q>Q_c$, the specific heat is always positive, implying the black holes are locally stable and no phase transition will take place. Secondly, both the $T-r_+$ curve and $T-S$ curve $f(R)$ AdS black holes are investigated and they exhibit Van der Vaals like behavior as the $P-v$ curve in the former research. Critical physical quantities are obtained and they are consistent with those derived from the specific heat analysis. We carry out numerical check of Maxwell equal area law for the cases $Q=0.2Q_c, 0.4Q_c, 0.6Q_c, 0.8Q_c$. The relative errors are amazingly small and can be negligible. So the Maxwell equal area law holds for $T-S$ curve of $f(R)$ black holes. Thirdly, we establish geometrothermodynamics for $f(R)$ AdS black hole to examine the phase structure. It is shown that the Legendre invariant scalar curvature $\mathfrak{R}$ would diverge exactly where the specific heat diverges. To summarize, the above three perspectives are consistent with each other, thus providing a unified picture which deepens the understanding of critical phenomena of $f(R)$ AdS black holes.}

\begin{document}
\maketitle
\flushbottom

\section{Introduction}
\label{sec:1}
To explain the cosmic acceleration, we usually seek help from two kinds of approach. One is to introduce the dark energy while the other is the so-called modified gravity theories. Among the latter approach, $f(R)$ gravity successfully mimic the cosmological history. It has various applications in both gravitation and cosmology. For nice reviews, see Ref.~\cite{Felice,Capozziello}. Black hole solutions in $f(R)$ gravity and their thermodynamics~\cite{Dombriz}-\cite{xiong5} are of interest. It is believed that the thermodynamics of $f(R)$ black holes distinguishes from that of black holes in Einstein gravity. Four-dimensional charged AdS black hole solution was obtained in the $R+f(R)$ gravity with constant curvature~\cite{Moon98}. Thermodynamic quantities, such as energy, entropy, heat capacity and Helmhotz free energy were discussed~\cite{Moon98}. $P-V$ criticality of four-dimensional $f(R)$ AdS black hole was also investigated~\cite{Chen}. It was proved that $f(R)$ gravity corrects the Gibbs free energy and the ratio $\rho_c$. Recently, we derived the analytic expression of the universal coexistence curve for $f(R)$ AdS black hole and obtained the explicit expression of the physical quantity describing the difference of the number densities of black hole molecules between the small and large black hole~\cite{xiong5}.

   In this paper, we would like to investigate the critical phenomena of $f(R)$ AdS black hole in the canonical ensemble. Specifically, we will probe in detail the behavior of the specific heat at constant charge. Although it was calculated in Ref.~\cite{Moon98}, its interesting properties that depend on the charge of black hole has not been covered. In fact, these properties provide another perspective other than $P-V$ criticality for one to understand the rich critical phenomena of $f(R)$ AdS black hole and certainly deserve to be further probed. Furthemore, we will study the critical phenomena from the perspective of $T-r_+$ curve and $T-S$ curve. It was first reported in literature that the $T-S$ curve (or $\beta-r_+$ curve) of RN-AdS black hole exhibits Van der Waal like behavior \cite{Chamblin1,Chamblin2} and Maxwell equal area law can be applied \cite{Spallucci}. So one may wonder whether $T-S$ curve of $f(R)$ AdS black hole shows similar behavior. One may also wonder whether the critical point of $f(R)$ AdS black hole obtained from $T-r_+$ curve and $T-S$ curve is consistent with that from the specific heat analysis.

   On the other hand, thermodynamic geometry serves as an elegant approach to probe the phase structure of black holes. Weinhold \cite{Weinhold} presented metric structure as $g_{i,j}^{W}=\partial_{i}\partial_{j}M(U,N^a)$ while the metric structures was proposed as $g_{i,j}^{R}=-\partial_{i}\partial_{j}S(U,N^a)$ by Ruppeiner \cite{Ruppeiner}. Moreover, Weinhold's metrics were found to be conformally connected to Ruppeiner's metrics through the map $dS^2_R=\frac{dS^2_W}{T}$\cite{Janyszek}. Recently, Quevedo et al. \cite{Quevedo2} proposed the so-called geometrothermodynamics, which successfully derives Legendre invariant metrics and reproduces the phase transition structures of various black holes \cite{Quevedo3}-\cite{Naderi}. In this paper, we would also like to construct the geometrothermodynamics for $f(R)$ AdS black hole. It is expected to provide not only an alternative perspective of the critical phenomena but also a tool to examine the phase structure of $f(R)$ AdS black hole.

    The organization of this paper is as follows. In Sec.~\ref{sec:2} we will have a brief review of the metric and relevant physical quantities of $f(R)$ AdS black hole. Phase transition will be investigated in canonical ensemble in Sec.~\ref {sec:3} while Van der Vaals like behavior of Hawking temperature and free energy will be studied and Maxwell equal area law will be numerically checked in Sec.~\ref{sec:4}. In Sec.~\ref {sec:5}, we will establish geometrothermodynamics for $f(R)$ AdS black hole to examine the phase structure. Conclusions will be drawn in Sec.~\ref {sec:6}.

\section{A brief review of charged AdS black holes in $f(R)$ gravity}
\label{sec:2}
  Charged AdS black hole solution in the $R+f(R)$ gravity with constant curvature scalar $R=R_0$ was obtained in Ref.~\cite{Moon98}. The metric reads
\begin{equation}
ds^2=-N(r)dt^2+\frac{dr^2}{N(r)}+r^2(d\theta^2+sin^2\theta d\phi^2),\label{1}
\end{equation}%
where
\begin{align}
N(r)&=1-\frac{2m}{r}+\frac{q^2}{br^2}-\frac{R_0}{12}r^2,\label{2}
\\
b&=1+f'(R_0).\label{3}
\end{align}%
Note that $b>0,R_0<0$. $m$ and $q$ are parameters related to the black hole ADM mass $M$ and the electric charge $Q$ respectively as follows~\cite{Moon98}
\begin{equation}
M=mb,\;\;\; Q=\frac{q}{\sqrt{b}}.\label{4}
\end{equation}%

When the curvature scalar is identified as~\cite{Moon98}
\begin{equation}
R_0=-\frac{12}{l^2}=4\Lambda,\label{5}
\end{equation}%
the above solution is asymptotically AdS.

The Hawking temperature, entropy and electric potential was reviewed in Ref.~\cite{Chen} as
\begin{align}
T&=\frac{N'(r_+)}{4\pi}=\frac{1}{4\pi r_+}(1-\frac{q^2}{br_+^2}-\frac{R_0r_+^2}{4}).\label{6}
\\
S&=\pi r_+^2b.\label{7}
\\
\Phi&=\frac{\sqrt{b}q}{r_+}.\label{8}
\end{align}

\section{Phase transition of $f(R)$ AdS black hole in canonical ensemble}
\label{sec:3}
To investigate the phase transition in canonical ensemble, one can observe the behavior of the specific heat at constant charge.

Substituting Eqs.(\ref{4}) and (\ref{7}) into Eq.(\ref{6}), one can reexpress the Hawking temperature as
\begin{equation}
T=\frac{4b\pi S-R_0 S^2-4b^2\pi^2Q^2}{16\pi^{3/2}\sqrt{bS^3}}.\label{9}
\end{equation}

\begin{figure*}
\centerline{\subfigure[]{\label{1a}
\includegraphics[width=8cm,height=6cm]{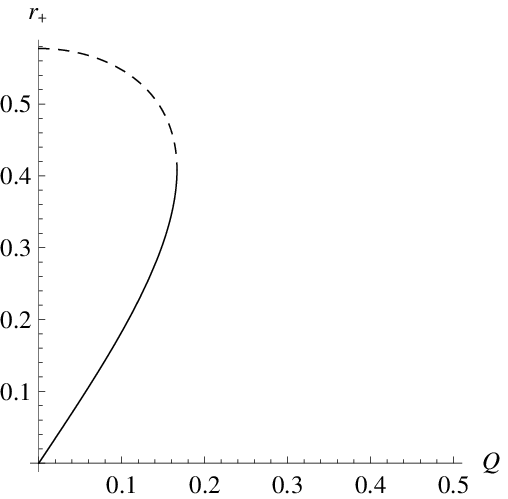}}
\subfigure[]{\label{1b}
\includegraphics[width=8cm,height=6cm]{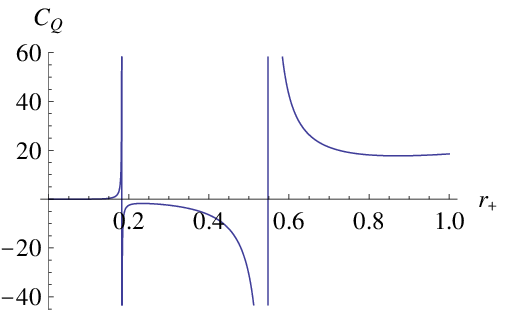}}}
\centerline{\subfigure[]{\label{1c}
\includegraphics[width=8cm,height=6cm]{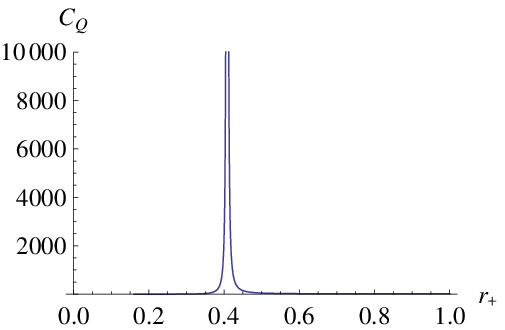}}
\subfigure[]{\label{1d}
\includegraphics[width=8cm,height=6cm]{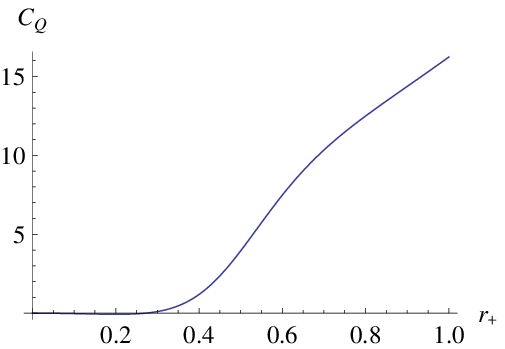}}}
 \caption{(a) The two roots of Eq.(\ref{11}) for $R_0=-12$ (b) The specific heat $C_Q$ vs. $r_+$ for $Q=0.1<Q_c, b=1.5, R_0=-12$ (c) The specific heat $C_Q$ vs. $r_+$ for $Q=\frac{1}{6}=Q_c, b=1.5, R_0=-12$ (d) The specific heat $C_Q$ vs. $r_+$ for $Q=0.3>Q_c, b=1.5, R_0=-12$} \label{fg1}
\end{figure*}

 The corresponding specific heat with charge fixed can be calculated as
\begin{equation}
C_Q=T(\frac{\partial S}{\partial
T})_Q=\frac{2S(4b^2\pi^2Q^2-4b\pi S+R_0S^2)}{-12b^2\pi^2Q^2+4b\pi S+R_0S^2}.\label{10}
\end{equation}
It is apparent that $C_q$ may diverge when $-12b^2\pi^2Q^2+4b\pi S+R_0S^2=0$. The denominator of $C_Q$ can be reexpressed as
\begin{equation}
\pi^2b^2(R_0r_+^4+4r_+^2-12Q^2)=0,\label{11}
\end{equation}
which can be solved as
\begin{equation}
r_+=\sqrt{2}\sqrt{-\frac{1}{R_0}\pm\frac{\sqrt{1+3R_0Q^2}}{R_0}}.\label{12}
\end{equation}
The solutions are depicted in Fig.\ref{1a}. These two solutions will degenerate into one when $1+3R_0Q^2=0$. This can also be observed in Fig.\ref{1a}. In other words, when
\begin{equation}
Q_c=\sqrt{\frac{-1}{3R_0}},\label{13}
\end{equation}
\begin{equation}
r_c=\sqrt{-\frac{2}{R_0}}.\label{14}
\end{equation}

The behavior of specific heat for the cases $Q<Q_c$, $Q=Q_c$, $Q>Q_c$ is shown in Fig.\ref{1b}, \ref{1c}, \ref{1d} respectively. The curve of specific heat for $Q<Q_c$ has two divergent points while that for $Q=Q_c$ has only one divergent point. Fig.\ref{1b} can be divided into three regions. Both the large radius region and the small radius region are thermodynamically stable with positive specific heat while the medium radius region is unstable with negative specific heat. So the phase transition takes place between small black hole and large black hole. However, when $Q>Q_c$, the specific heat is always positive, implying the black holes are locally stable and no phase transition will take place.

%%%%%%%%%%%%%%%%%%%%%%%%%%%%%%%%%%%%%%%%%%%%%%%%%%%%%%%%%%%%%%%%%%%%%%%%%%%%%
%%%%%%%%%%%%%%%%%%%%%%%%%%%%%%%%%%%%%%%%%%%%%%%%%%%%%%%%%%%%%%%%%%%%%%%%%%%%%%%%

Besides the specific heat, the inverse of the isothermal compressibility deserves to be studied. It is defined as
\begin{equation}
\kappa _T^{-1}=Q(\frac{\partial\Phi}{\partial Q})_T.\label{15}
\end{equation}
which can be calculated using the thermodynamic identity relation as follow
\begin{equation}
(\frac{\partial\Phi}{\partial T})_Q(\frac{\partial T}{\partial
Q})_\Phi(\frac{\partial Q}{\partial \Phi})_T=-1.\label{16}
\end{equation}
Then the explicit form of $\kappa _T^{-1}$ for $f(R)$ AdS black holes can be derived as
\begin{equation}
\kappa
_T^{-1}=\frac{bQ(R_0r_+^4+4r_+^2-4Q^2)}{r_+(R_0r_+^4+4r_+^2-12Q^2)}.\label{17}
\end{equation}

%%%%%%%%%%%%%%%%%%%%%%%%%%%%%%%%%%%%%%%%%%%%%%%%%%%%%%%%%%%%%%%%%%%%%%%%%%%%%
\begin{figure}
\includegraphics[width=8cm,height=6cm]{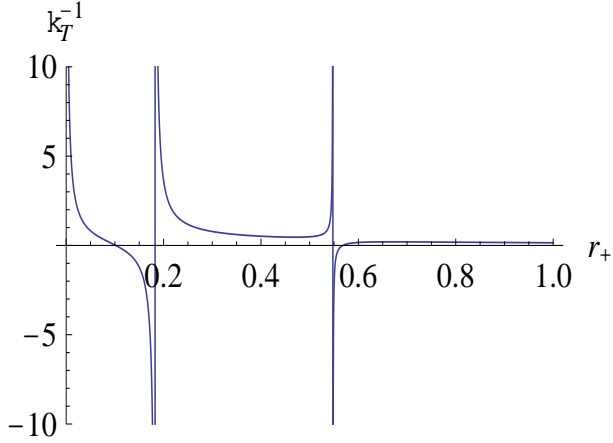}
 \caption{The inverse of the isothermal compressibility $\kappa _T^{-1}$ vs. $r_+$ for $Q=0.1<Q_c, b=1.5, R_0=-12$}
\label{fg2}
\end{figure}
%%%%%%%%%%%%%%%%%%%%%%%%%%%%%%%%%%%%%%%%%%%%%%%%%%%%%%%%%%%%%%%%%%%%%%%%%%%%%%%%
Comparing Eq.(\ref{17}) with Eq.(\ref{11}), one can see clearly that $\kappa _T^{-1}$ shares the same term
$R_0r_+^4+4r_+^2-12Q^2$ in its denominator, implying that it may also diverge where the specific heat does. The divergent behavior of $\kappa _T^{-1}$ is shown in Fig.\ref{fg2}.

\section{Van der Vaals like behavior of Hawking temperature and Maxwell equal area law}
\label{sec:4}

It was shown in Ref.~\cite{Chen} that when the cosmological constant is identified as thermodynamic pressure, $P-v$ graph exhibits Van der Vaals like behavior. Here, we will show that Hawking temperature of charged AdS black holes in $f(R)$ gravity also possess the interesting Van der Vaals like property.

The possible critical point can be determined through
\begin{eqnarray}
\left(\frac{\partial T}{\partial r}\right)_{Q=Q_c, r=r_c}&=&0,\label{18}
\\
\left(\frac{\partial^2 T}{\partial r^2}\right)_{Q=Q_c, r=r_c}&=&0.\label{19}
\end{eqnarray}%

Solving the above equations, one can obtain
\begin{equation}
r_c=\sqrt{\frac{-2}{R_0}},\;\;\; Q_c=\sqrt{\frac{-1}{3R_0}}.\label{20}
\end{equation}%

%%%%%%%%%%%%%%%%%%%%%%%%%%%%%%%%%%%%%%%%%%%%%%%%%%%%%%%%%%%%%%%%%%%%%%%%%%%%%
\begin{figure*}
\centerline{\subfigure[]{\label{3a}
\includegraphics[width=8cm,height=6cm]{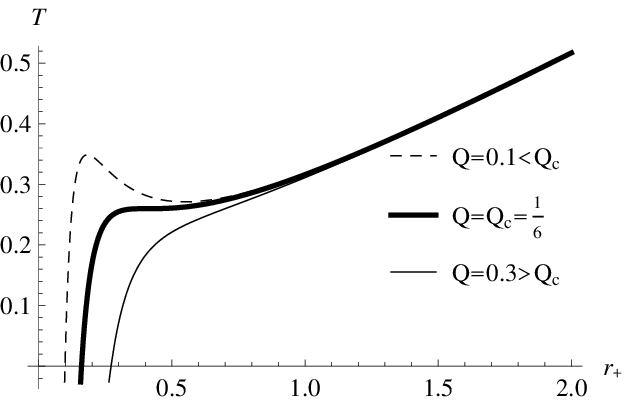}}
\subfigure[]{\label{3b}
\includegraphics[width=8cm,height=6cm]{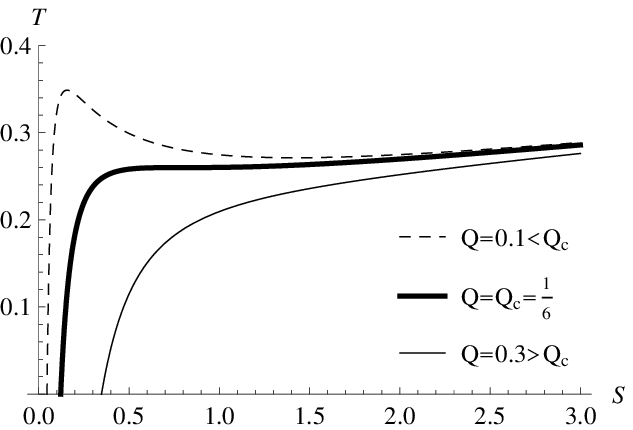}}}
 \caption{(a) $T$ vs. $r_+$ for $b=1.5, R_0=-12$ (b) $T$ vs. $S$ for $b=1.5, R_0=-12$} \label{fg3}
\end{figure*}
%%%%%%%%%%%%%%%%%%%%%%%%%%%%%%%%%%%%%%%%%%%%%%%%%%%%%%%%%%%%%%%%%%%%%%%%%%%%%%%%

It can be witnessed from Eq.(\ref{20}) that the above critical quantities only depend on $R_0$. Fig. \ref{3a} shows the curves of Hawking temperature corresponding to different values of charge $Q$. Note that we choose $R_0=-12, b=1.5$ in Fig. \ref{3a}. When $Q<Q_c$, the curve can be divided into three branches. The slope of the large radius branch and the small radius branch are both positive while the slope of the medium radius branch is negative. When $Q>Q_c$, the Hawking temperature increases monotonically. This phenomenon is analogous to the van der Waals liquid-gas system.

$T-S$ curve, depicted in Fig. \ref{3b}, also exhibits similar behavior. The corresponding critical point can be defined as
\begin{align}
\left(\frac{\partial T}{\partial S}\right)_{Q=Q_c, S=S_c}&=0,\label{21}
\\
\left(\frac{\partial^2 T}{\partial S^2}\right)_{Q=Q_c, S=S_c}&=0.\label{22}
\end{align}%
Substituting Eq.(\ref{9}) into the above two equations, one can obtain
\begin{equation}
S_c=\frac{-2b\pi}{R_0},\;\;\; Q_c=\sqrt{\frac{-1}{3R_0}}.\label{23}
\end{equation}%

Comparing Eq.(\ref{23}) with Eqs.(\ref{13}), (\ref{14}) and (\ref{20}), one can find that the $T-r_+$ curve, the $T-S$ curve and the specific heat analysis are consistent with each other. In Sec.~\ref {sec:3} we have shown that both the large radius branch and the small radius branch are stable with positive specific heat while the medium radius branch is unstable with negative specific heat. As argued in Ref. \cite{Spallucci}, one can remove the unstable branch in $T-S$ curve with a bar vertical to the temperature axis $T=T_*$ and examine the Maxwell equal area law as follow
\begin{equation}
T_*(S_3-S_1)=\int^{S_3}_{S_1}TdS,\label{24}
\end{equation}%
where $S_1$, $S_2$, $S_3$ denote the three values of the entropy corresponding to $T=T_*$. And We assume that $S_1<S_2<S_3$.

To facilitate the numerical check of Maxwell equal area law, we would like to first probe the behavior of free energy, which can be derived as
\begin{equation}
F=M-TS=\frac{R_0S^2+12\pi b(3b\pi Q^2+S)}{48\pi^{3/2}\sqrt{bS}}.\label{25}
\end{equation}%

\begin{figure*}
\centerline{\subfigure[]{\label{4a}
\includegraphics[width=8cm,height=6cm]{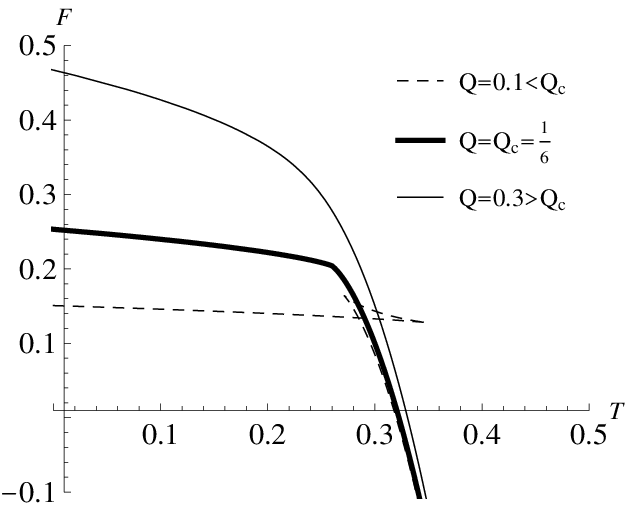}}
\subfigure[]{\label{4b}
\includegraphics[width=8cm,height=6cm]{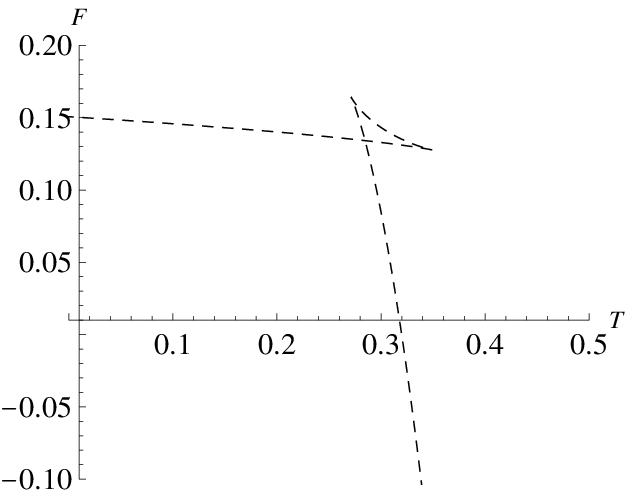}}}
 \caption{(a) $F$ vs. $T$ for $b=1.5, R_0=-12$ (b) $F$ vs. $T$ for $b=1.5, R_0=-12, Q=0.1<Q_c$} \label{fg4}
\end{figure*}

The behavior of free energy is plotted in Fig. \ref{4a} for different choices of electric charge. It is shown that the classical swallow tails characteristic of first order phase transition appears in the case $Q<Q_c$. To witness this phenomenon more clearly, we plot this case independently in Fig. \ref{4b}.
%%%%%%%%%%%%%%%%%%%%%%%%%%%%%%%%%%%%%%%%%%%%%%%%%%%%%%%%%%%%%%%%%%%%%%%%%%%%%
%%%%%%%%%%%%%%%%%%%%%%%%%%%%%%%%%%%%%%%%%%%%%%%%%%%%%%%%%%%%%%%%%%%%%%%%%%%%%%%%

We carry out numerical check of Maxwell equal area law for the cases $Q=0.2Q_c, 0.4Q_c,$ $0.6Q_c, 0.8Q_c$. We utilize the free energy analysis to determine
$T_*$ and calculate respectively the values of the left-hand side and right-hand side of Eq.(\ref{24}). As shown in Table. \ref{tb1}, the relative errors for the four cases are amazingly small and can be negligible. So the Maxwell equal area law holds for $T-S$ curve of $f(R)$ black holes.

\begin{table}[!h]
\tabcolsep 0pt
\caption{Numerical check of Maxwell equal area law for $b=1.5, R_0=-12$}
\vspace*{-12pt}
\begin{center}
\def\temptablewidth{1\textwidth}
{\rule{\temptablewidth}{2pt}}
\begin{tabular*}{\temptablewidth}{@{\extracolsep{\fill}}cccccccc}
$Q$ & $T_*$ & $S_1$ &$S_2$ &$S_3$ &$T_*(S_3-S_1)$ & $\int^{S_3}_{S_1}TdS$ & relative error \\   \hline
    $0.2Q_c$    & 0.30751656     & 0.00605049       &      0.588755 & 4.07802 & 1.25219806 & 1.25219807 & $7.98596\times10^{-9}$ \\
     $0.4Q_c$  & 0.29633040     & $0.02879995$        &        0.646256 & 3.42695  & 1.00697516 & 1.00697508 & $7.94459\times10^{-8}$ \\
 $0.6Q_c$     &  0.28470500     & $0.08085180$        &        0.697672 & 2.74658 & 0.75894615 & 0.75894619 & $5.27047\times10^{-8}$ \\
   $0.8Q_c$    & 0.27258400     & $0.19720287$        &        0.743896 & 2.00189 & 0.49192884 & 0.49192938 & $1.09772\times10^{-6}$
       \end{tabular*}
       {\rule{\temptablewidth}{2pt}}
       \end{center}
       \label{tb1}
       \end{table}

\section{Geometrothermodynamics in canonical ensemble}
\label{sec:5}
    According to geometrothermodynamics proposed by Quevedo, the non-degenerate metric $G$ and the thermodynamic metric $g$ can be written as
          follows~\cite{Quevedo7}
\begin{align}
G&=(d\phi-\delta_{ab}I^adE^b)^2+(\delta_{ab}E^aI^b)(\eta_{cd}dE^cdI^d),\label{26}
\\
g&=\varphi^*(G)=(E^c\frac{\partial \phi}{\partial
E^c})(\eta_{ab}\delta^{bc}\frac{\partial^2\phi}{\partial E^c
\partial E^d}dE^adE^d),\label{27}
\end{align}%
where $\eta_{ab}=diag(-1,\cdots,1)$.

   Choosing $M$ to be the thermodynamic potential and $S,Q$ to be the
extensive variables, one can construct a five-dimensional thermodynamic phase
space coordinated by the set of independent
coordinates\{$M, S, Q, T, \Phi$\}. Then the non-degenerate metric $G$ for the $f(R)$ AdS black holes can be written as
\begin{equation}
G=(dM-TdS-\Phi dQ)^2+(TS+\Phi Q)(-dSdT+dQd\Phi).\label{28}
\end{equation}%
The space of thermodynamic equilibrium states can be induced via introducing the map
\begin{equation}
\varphi:\{S,Q\}\mapsto\{M(S,Q),S,Q,\frac{\partial M}{\partial
S},\frac{\partial M}{\partial Q}\}.\label{29}
\end{equation}%

Utilizing Eq.(\ref{27}), the thermodynamic metric $g$ can be obtained as
\begin{equation}
g=(S\frac{\partial M}{\partial S}+Q\frac{\partial M}{\partial
Q})(-\frac{\partial^2M}{\partial S^2}dS^2+\frac{\partial^2
M}{\partial Q^2}dQ^2).\label{30}
\end{equation}%
The relevant quantities can be obtained utilizing Eqs.(\ref{2}), (\ref{4}) and (\ref{7})
\begin{align}
\frac{\partial M}{\partial S}&=\frac{4b\pi S-R_0 S^2-4b^2\pi^2Q^2}{16\pi^{3/2}\sqrt{bS^3}},\label{31}
\\
\frac{\partial M}{\partial
Q}&=\frac{b^{3/2}\sqrt{\pi}Q}{\sqrt{S}},\label{32}
\\
\frac{\partial^2 M}{\partial
S^2}&=\frac{12b^2\pi^2Q^2-4b\pi S-R_0S^2}{32\pi^{3/2}\sqrt{bS^5}},\label{33}
\\
\frac{\partial^2 M}{\partial Q^2}&=\frac{b^{3/2}\sqrt{\pi}}{\sqrt{S}}.\label{34}
\end{align}%

Comparing Eqs.(\ref{31}), (\ref{32}) with Eqs.(\ref{9}), (\ref{8}), one can find that
 \begin{equation}
\frac{\partial M}{\partial S}=T,\;\;\; \frac{\partial M}{\partial Q}=\Phi,\label{35}
\end{equation}
suggesting that the first law of black hole thermodynamics $dM=TdS+\Phi dQ$ holds for $f(R)$ AdS black holes.

Substituting Eqs.(\ref{31})-(\ref{34}) into Eq.(\ref{30}), the component of the thermodynamic metric $g$ can be derived as
\begin{align}
g_{SS}&=-\frac{9b^3\pi Q^4}{32S^3}+\frac{b(2+3Q^2R_0)}{64\pi S}-\frac{R_0^2S}{512b\pi^3},\label{36}
\\
g_{QQ}&=\frac{b^2}{4}+\frac{3b^3\pi Q^2}{4S}-\frac{bR_0S}{16\pi}.\label{37}
\end{align}%

Programming with Mathematica 9.0, one can obtain the scalar curvature as
\begin{equation}
\mathfrak{R}=\frac{A(S,Q)}{(R_0S^2-4b\pi S-12b^2\pi^2Q^2)^3(R_0S^2+4b\pi S-12b^2\pi^2Q^2)^2},\label{38}
\end{equation}%
where
\begin{align}
A(S,Q)&=256b\pi^3S^2\times\big[1152b^5\pi^5Q^4+144b^4\pi^4Q^2S\times(7Q^2R_0-8)-240b^3\pi^3Q^2R_0S^2
\nonumber
\\
&\;-8b^2\pi^2R_0S^3\times(3Q^2R_0-4)-4b\pi R_0^2S^4-5R_0^3S^5\big].\label{39}
\end{align}%

\begin{figure}
\includegraphics[width=8cm,height=6cm]{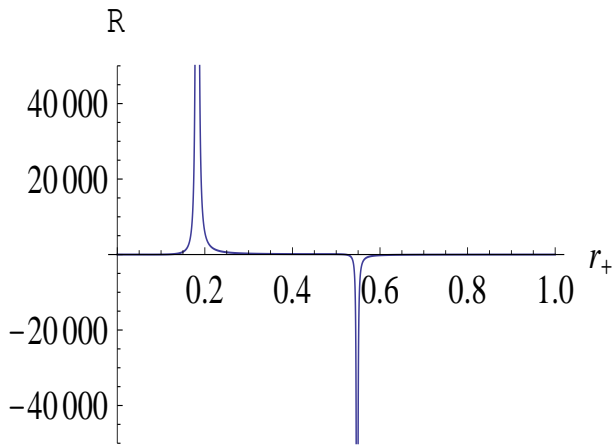}
 \caption{Thermodynamic scalar curvature $\mathfrak{R}$ vs. $r_+$ for $Q=0.1<Q_c, b=1.5, R_0=-12$}
\label{fg5}
\end{figure}

Note that $R_0<0, S>0, b>0$, one can deduce that $R_0S^2-4b\pi S-12b^2\pi^2Q^2<0$. So the Legendre invariant scalar curvature $\mathfrak{R}$ would diverge when $R_0S^2+4b\pi S-12b^2\pi^2Q^2=0$. Comparing it with Eq.(\ref{10}), it is exactly the same as the denominator of the specific heat $C_Q$, implying that it would diverge where the specific heat diverges. Fig.~\ref{fg4} presents an intuitive understanding of the Legendre invariant scalar curvature $\mathfrak{R}$. Comparing Fig.~\ref{fg5} with Figure~\ref{1b}, one can conclude that the Legendre invariant metric constructed in geometrothermodynamics correctly produces phase transition structure of $f(R)$ AdS black holes.

%%%%%%%%%%%%%%%%%%%%%%%%%%%%%%%%%%%%%%%%%%%%%%%%%%%%%%%%%%%%%%%%%%%%%%%%%%%%%
%%%%%%%%%%%%%%%%%%%%%%%%%%%%%%%%%%%%%%%%%%%%%%%%%%%%%%%%%%%%%%%%%%%%%%%%%%%%%%%%

\section{Conclusions and Discussions}
\label{sec:6}
  In this paper, we probe the critical phenomena and geometrothermodynamics of $f(R)$ AdS black holes. Firstly, we investigate the phase transition in canonical ensemble. We study in detail the specific heat $C_Q$ when the charge of $f(R)$ AdS black hole is fixed and derive the explicit solutions corresponding to the divergence of $C_Q$. The two solutions merge into one when the condition $Q_c=\sqrt{\frac{-1}{3R_0}}$ is satisfied. The cases $Q<Q_c$, $Q=Q_c$, $Q>Q_c$ are shown intuitively. The curve of specific heat for $Q<Q_c$ has two divergent points and can be divided into three regions. Both the large radius region and the small radius region are thermodynamically stable with positive specific heat while the medium radius region is unstable with negative specific heat. So the phase transition takes place between small black hole and large black hole. However, when $Q>Q_c$, the specific heat is always positive, implying the black holes are locally stable and no phase transition will take place.

  Secondly, we study the behavior of Hawking temperature. Both the $T-r_+$ curve and $T-S$ curve exhibit the Van der Vaals like behavior as $P-v$ curve in the former literature \cite{Chen}. Critical physical quantities are obtained and they are consistent with those derived from the specific heat analysis. When $Q<Q_c$, the curves can be divided into three branches. The slope of the large radius branch and the small radius branch are both positive while that of the medium radius branch is negative. When $Q>Q_c$, the Hawking temperature increases monotonically. The behavior of free energy is plotted and it is shown that the classical swallow tails characteristic of first order phase transition appears in the case $Q<Q_c$. We carry out numerical check of Maxwell equal area law for the cases $Q=0.2Q_c, 0.4Q_c, 0.6Q_c, 0.8Q_c$. The relative errors are amazingly small and can be negligible. So the Maxwell equal area law holds for $T-S$ curve of $f(R)$ black holes.

   Thirdly, we establish geometrothermodynamics for $f(R)$ AdS black hole to examine the phase structure. Choosing $M$ to be the thermodynamic potential and $S,Q$ to be the extensive variables, we construct a five-dimensional thermodynamic phase space coordinated by the set of independent coordinates\{$M, S, Q, T, \Phi$\}. We derive both the non-degenerate metric $G$ and the thermodynamic metric $g$ for the $f(R)$ AdS black holes. It is proved that the first law of black hole thermodynamics $dM=TdS+\Phi dQ$ holds for $f(R)$ AdS black holes. Programming with Mathematica 9.0, we obtain the explicit expression of scalar curvature $\mathfrak{R}$. It is shown that the Legendre invariant scalar curvature $\mathfrak{R}$ would diverge when $R_0S^2+4b\pi S-12b^2\pi^2Q^2=0$. It is exactly where the specific heat diverges. So the Legendre invariant metric constructed in geometrothermodynamics correctly produces phase transition structure of $f(R)$ AdS black holes.

   Recently, the isocharges in the entanglement entropy-temperature plane was investigated \cite{Johnson}. It was shown that both the critical temperature and critical exponent are exactly the same as the case of entropy-temperature plane \cite{Johnson}. This intriguing finding has attracted more and more attention \cite{Caceres}-\cite{zengxiaoxiong3}. Especially, Ref. \cite{Nguyen} further proved that the Van der Waals behavior on the entanglement entropy-temperature plane also obeys the equal area law. It would be interesting to probe under the background of $f(R)$ AdS black hole and investigate the behavior of its entanglement entropy. It is expected to provide us with a holographic perspective of critical phenomena of $f(R)$ AdS black hole. We will return to this issue in the near future.

   On the other hand, it will also be interesting to probe the possible observational signatures to be detected for black hole phase transitions. In a recent interesting paper \cite{liuyunqi}, the authors calculated the quasinormal modes of massless scalar perturbations around small and large four-dimensional RN-AdS black holes. A dramatic change was found in the slopes of quasinormal frequencies in small and large black holes near the critical point. Their results further confirmed the idea that the quasinormal mode can be a dynamic probe of the thermodynamic phase transition that was proposed in former literatures \cite{jingjiliang}-\cite{hexi}. Inspired by these research, we believe that it would be a promising direction to probe in $f(R)$ AdS black holes the relation between the phase transition and the stability against perturbation. Concerning the stability $f(R)$ black holes of under perturbation, there has been nice paper dealing with $f(R)$ (Schwarzschild) black hole \cite{Myung1}, $f(R)$-AdS (Schwarzschild-AdS) black holes \cite{Myung2}, rotating black hole in a limited form of $f(R)$ gravity \cite{Myung3}, Kerr black holes in $f(R)$ gravity \cite{Myung4}. Research on stability of charged $f(R)$ AdS black holes against linear perturbation and its relation with phase transition is called for. And it would certainly help broaden the perspective in this paper.

\acknowledgments We would like to express our sincere gratitude to the anonymous referee whose insightful suggestion has improved the quality of this paper greatly. This research is supported by Guangdong Natural Science Foundation (Grant No.2015A030313789) and Department of Education of Guangdong Province of China(Grant No.2014KQNCX191). It is also supported by \textquotedblleft Thousand Hundred Ten\textquotedblright \,Project of Guangdong Province.


\begin{thebibliography}{99}



\bibitem{Felice}
A. De Felice and S. Tsujikawa, $f(R)$ theories, Living Rev. Rel. 13 (2010) 3

\bibitem{Capozziello}
S. Capozziello and M. De Laurentis, Extended Theories of Gravity, Phys. Rept. 509 (2011) 167-321

\bibitem{Dombriz}
A. de la C. Dombriz, A. Dobado and A.L. Maroto, Black Holes in $f(R)$ theories, Phys. Rev. D 80(2009)124011



\bibitem{Moon98}
T. Moon, Y. S. Myung and E. J. Son, $f(R)$ Black holes, Gen. Rel. Grav. 43(2011)3079-3098.


\bibitem{Larranaga}
A. Larranaga, A Rotating Charged Black Hole Solution in $f(R)$ Gravity, Pramana 78(2012)697-703

\bibitem{Cembranos}
J. A. R. Cembranos, A. de la C. Dombriz and P. J. Romero, Kerr-Newman black holes in f(R) theories, Int. J. Geom. Meth. Mod. Phys. 11(2014)1450001

\bibitem{Sheykhi}
A. Sheykhi, Higher-dimensional charged $f(R)$ black holes, Phys. Rev. D 86(2012)024013



\bibitem{Sebastiani}
L. Sebastiani and S. Zerbini, Static Spherically Symmetric Solutions in $F(R)$ Gravity, Eur. Phys. J. C 71(2011)1591





\bibitem{Hendi3}
S. H. Hendi, The Relation between $F(R)$ gravity and Einstein-conformally invariant Maxwell source, Phys. Lett. B 690(2010)220-223

\bibitem{Hendi4}
S. H. Hendi and D. Momeni, Black hole solutions in $F(R)$ gravity with conformal anomaly, Eur. Phys. J. C71(2011)1823

\bibitem{Olmo}
G. J. Olmo and D. R. Garcia, Palatini $f(R)$ Black Holes in Nonlinear Electrodynamics, Phys. Rev. D 84(2011)124059

\bibitem{Mazharimousavi}
S. H. Mazharimousavi and M. Halilsoy, Black hole solutions in $f(R)$ gravity coupled with non-linear Yang-Mills field, Phys. Rev. D 84(2011)064032

\bibitem{Myung1}
Y. S. Myung, T. Moon and E. J. Son, Stability of $f(R)$ black holes, Phys. Rev. D 83 (2011) 124009

\bibitem{Myung2}
T. Moon, Y. S. Myung and E. J. Son, Stability analysis of $f(R)$-AdS black holes, Eur. Phys. J. C 71 (2011) 1777

\bibitem{Myung3}
Y. S. Myung, Instability of rotating black hole in a limited form of $f(R)$ gravity, Phys. Rev. D 84 (2011) 024048

\bibitem{Myung4}
Y. S. Myung, Instability of a Kerr black hole in $f(R)$ gravity, Phys. Rev. D 88 (2013)104017




\bibitem{Chen}
S. Chen, X. Liu, C. Liu and J. Jing, $P-V$ criticality of AdS black hole in $f(R)$ gravity, Chin. Phys. Lett. 30(2013)060401

\bibitem{xiong5}
J. X. Mo and G. Q. Li, Coexistence curves and molecule number densities of AdS black holes in the reduced parameter space, Phys. Rev. D92 (2015)024055


\bibitem{Chamblin1}
 A. Chamblin, R. Emparan, C.V. Johnson and R.C. Myers, Charged AdS Black Holes and Catastrophic Holography, Phys. Rev. D 60(1999)064018

\bibitem{Chamblin2}
A. Chamblin, R. Emparan, C. V. Johnson and R.C. Myers, Holography, Thermodynamics and Fluctuations of Charged AdS Black Holes, Phys. Rev. D 60(1999)104026

\bibitem{Spallucci}
E. Spallucci and A. Smailagic, Maxwell's equal area law for charged Anti-de Sitter black holes, Phys. Lett. B 723(2013)436-441

\bibitem{Weinhold}
F. Weinhold, Metric geometry of equilibrium thermodynamics, Chem. Phys. 63(1975)2479


\bibitem{Ruppeiner}
G. Ruppeiner, A Riemannian geometric model, Phys. Rev. A 20(1979)1608

\bibitem{Janyszek}
H. Janyszek, R. Mrugala, Geometrical structure of the state space inclassical statistical and phenomenological thermodynamics, Rep. Math. Phys. 27 (1989)145




\bibitem{Quevedo2}
H. Quevedo, Geometrothermodynamics, J. Math. Phys. 48(2007)013506

\bibitem{Quevedo3}
H. Quevedo, Geometrothermodynamics of black holes, Gen. Rel. Grav. 40(2008)971-984


\bibitem{Quevedo4}
H. Quevedo and A. Sanchez, Geometrothermodynamics of asymptotically anti-de Sitter black holes, JHEP 0809(2008)034



\bibitem{Alvarez}
J. L. Alvarez, H. Quevedo and A. Sanchez, Unified geometric description of black hole thermodynamics, Phys. Rev. D 77(2008)084004


\bibitem{Quevedo5}
H. Quevedo and A. Sanchez, Geometrothermodynamics of black holes in two dimensions, Phys. Rev. D 79(2009)087504



\bibitem{Quevedo6}
H. Quevedo and  A. Sanchez, Geometric description of BTZ black holes thermodynamics, Phys. Rev. D 79(2009)024012



\bibitem{Akbar}
M. Akbar, H. Quevedo, K. Saifullah and A. Sanchez, S. Taj, Thermodynamic Geometry Of Charged Rotating BTZ Black Holes, Phys. Rev. D 83 (2011)084031


\bibitem{Quevedo7}
 H. Quevedo, A. Sanchez, S. Taj and A. Vazquez, Phase transitions in geometrothermodynamics, Gen. Rel. Grav 43(2011)1153-1165

\bibitem{Quevedo8}
H. Quevedo, A. Sanchez, S. Taj and A. Vazquez, Geometrothermodynamics in Horava-Lifshitz gravity,  J. Phys. A45 (2012) 055211

\bibitem{Quevedo9}
S. Taj and H. Quevedo, Geometrothermodynamics of five dimensional black holes in Einstein-Gauss-Bonnet-theory, Gen. Rel. Grav. 44 (2012) 1489-1523

\bibitem{Quevedo10}
A. Bravetti, D. Momeni, R. Myrzakulov and H. Quevedo, Geometrothermodynamics of higher dimensional black holes, Gen. Rel. Grav. 45 (2013) 1603-1617

\bibitem{Quevedo11}
H. Quevedo, M. N. Quevedo, A. Sanchez and S. Taj, On the ensemble dependence in black hole geometrothermodynamics, Phys. Scripta 8 (2014) 084007

\bibitem{Quevedo12}
H. Quevedo, M. N. Quevedo, A. Sanchez and S. Taj, Geometrothermodynamics of phantom AdS black holes, arXiv:1601.07120

\bibitem{Janke}
W. Janke, D.A. Johnston and R. Kenna, Geometrothermodynamics of the Kehagias-Sfetsos Black Hole, J. Phys. A 43 (2010)425206



\bibitem{Larranaga2}
A. Larranaga and A. Cardenas, Geometric Thermodynamics of Schwarzschild-AdS black hole with a Cosmological Constant as State Variable, J. Korean Phys. Soc. 60 (2012) 987-992

\bibitem{Rodrigues}
M. E. Rodrigues and Z. A. A. Oporto, Thermodynamics of phantom black holes in Einstein-Maxwell-Dilaton theory, Phys. Rev. D 85 (2012) 104022

\bibitem{Hanyiwen}
Y. W. Han and G. Chen, Thermodynamics, geometrothermodynamics and critical behavior of (2+1)-dimensional black hole, Phys. Lett. B 714(2012)127-130

\bibitem{Rodrigues2}
M. Azreg-A\"{i}nou and M. E. Rodrigues, Thermodynamical, geometrical and Poincar¨¦ methods for charged black holes in presence of quintessence, JHEP 1309 (2013) 146


\bibitem{Bravetti}
A. Bravetti, D. Momeni, R. Myrzakulov and A. Altaibayeva, Geometrothermodynamics of Myers-Perry black holes, Adv. High Energy Phys. 2013 (2013) 549808


\bibitem{jiexiong1}
J. X. Mo, X. X. Zeng, G. Q. Li, X. Jiang and W. B. Liu, A unified phase transition picture of the charged topological black hole in Horava-Lifshitz gravity, JHEP 1310 (2013) 056


\bibitem{jiexiong2}
J. X. Mo and W. B. Liu, Phase transitions, geometrothermodynamics and critical exponents of black holes with conformal anomaly, Adv. High Energy Phys. 2014 (2014) 739454



\bibitem{mamengsen}
M. S. Ma, Thermodynamics and phase transition of black hole in an asymptotically safe gravity, Phys. Lett. B735 (2014) 45-50


\bibitem{Suresh1}
J. Suresh, R. Tharanath, N. Varghese and V. C. Kuriakose, The thermodynamics and thermodynamic geometry of the Park black hole, Eur. Phys. J. C74 (2014) 2819



\bibitem{Suresh2}
R. Tharanath, J. Suresh and V. C. Kuriakose, Phase transitions and Geometrothermodynamics of Regular black holes, Gen.Rel.Grav. 47 (2015)46

\bibitem{Suresh3}
J. Suresh, R. Tharanath and V. C. Kuriakose, A unified thermodynamic picture of Ho\v{r}ava-Lifshitz black hole in arbitrary space time,  JHEP 1501 (2015) 019

\bibitem{zhangjialin1}
J. L. Zhang, R. G. Cai and H. Yu, Phase transition and thermodynamical geometry for Schwarzschild AdS black hole in Ad$S_5 ¡Á S^5$ spacetime,  JHEP 1502 (2015) 143

\bibitem{zhangjialin2}
J. L. Zhang, R. G. Cai and H. Yu, Phase transition and thermodynamical geometry of Reissner-Nordstr?m-AdS black holes in extended phase space, Phys. Rev. D91 (2015)044028

\bibitem{Panahiyan}
S. H. Hendi, S. Panahiyan, B. Eslam Panah, Geometrical method for thermal instability of nonlinearly charged BTZ Black Holes, Adv. High Energy Phys. 2015 (2015) 743086

\bibitem{Naderi}
S. H. Hendi and R. Naderi, Geometrothermodynamics of black holes in Lovelock gravity with a nonlinear electrodynamics, Phys. Rev. D91 (2015) 2, 024007




\bibitem{Johnson}
C. V. Johnson, Large $N$ Phase Transitions, Finite Volume, and Entanglement Entropy, JHEP 1403 (2014) 047

\bibitem{Caceres}
E. Caceres, P. H. Nguyen and J. F. Pedraza, Holographic entanglement entropy and the extended phase structure of STU black holes, JHEP 1509 (2015) 184

\bibitem{Nguyen}
P. H. Nguyen, An equal area law for holographic entanglement entropy of the AdS-RN black hole, JHEP 1512 (2015) 139

\bibitem{zengxiaoxiong1}
X. X. Zeng, H. Zhang and L. F. Li, Phase transition of holographic entanglement entropy in massive gravity, arXiv:1511.00383


\bibitem{Dey}
A. Dey, S. Mahapatra, T. Sarkar, Thermodynamics and Entanglement Entropy with Weyl Corrections, arXiv:1512.07117

\bibitem{zengxiaoxiong2}
X. X. Zeng and L. F. Li, Van der Waals phase transition in the framework of holography, arXiv:1512.08855



\bibitem{zengxiaoxiong3}
X. X. Zeng, X. M. Liu and L. F. Li, Phase structure of the Born-Infeld-anti-de Sitter black holes probed by non-local observables, arXiv:1601.01160

\bibitem{liuyunqi}
Y. Liu, D. C. Zou, and B. Wang,	Signature of the Van der Waals like small-large charged AdS black hole phase transition in quasinormal modes,  JHEP 1409 (2014) 179

\bibitem{jingjiliang}
J. Jing and Q. Pan, Quasinormal modes and second order thermodynamic phase transition
for Reissner-Nordstrom black hole, Phys. Lett. B 660 (2008) 13



\bibitem{wangbin}
J. Shen, B. Wang, C. Y. Lin, R. G. Cai and R. K. Su, The phase transition and the
Quasi-Normal Modes of black Holes, JHEP 0707 (2007) 037

\bibitem{hexi}
X. He, B. Wang, R. G. Cai and C. Y. Lin, Signature of the black hole phase transition in
quasinormal modes, Phys. Lett. B 688 (2010) 230

\end{thebibliography}
\end{document}